\documentclass[12pt]{revtex4}
\usepackage[pdftex]{graphicx}
\def \beq {\begin{equation}}
\def \eeq {\end{equation}}
\def \ba {\begin{eqnarray}}
\def \ea {\end{eqnarray}}

\begin{document}

\title{Tensor supercurrent in QCD}

\author{Alexander S. Gorsky$^{1,2}$ and Dmitri E. Kharzeev $^{3,4}$}

\affiliation{
$^1$ Institute for Information Transmission Problems of the Russian Academy of Sciences, Moscow,
Russia \\
$^2$ Moscow Institute of Physics and Technology, Dolgoprudny 141700, Russia \\
$^3$   Department  of  Physics  and  Astronomy,
Stony  Brook  University,  New  York  11794,  USA
            \\
$^4$ Physics  Department  and  RIKEN-BNL  Research  Center,
Brookhaven  National  Laboratory,  Upton,  NY  11973,  USA 
}

\begin{abstract}

An external Abelian magnetic field excites in the QCD vacuum a tensor supercurrent that represents the tensor polarization of the chiral condensate. This tensor supercurrent can be deduced  from the chiral lagrangian in the presence of anomalies; a similar tensor supercurrent emerges in rotating systems at finite chemical potential. We discuss the microscopic origin of this supercurrent  and argue that it screens the instanton--anti-instanton $I\bar{I}$ molecules  in the QCD vacuum, similarly to the vector supercurrent  screening Abrikosov vortices in a superconductor. A number of possible experimental manifestations of the tensor supercurrent are discussed:
{\it i}) spin alignment of axial-vector and vector mesons in heavy ion collisions; {\it ii}) tensor charge of the nucleon; {\it iii}) transversity of quark distributions in polarized nucleons.

% including electrically polarized pions and spin-aligned vector mesons. The tensor supercurrent can also be viewed as a 2-form current in so-called 1-form hydrodynamics, with the density of strings proportional to the chiral condensate. %Possible experimental manifestations of the tensor supercurrent include the spin alignment of vector mesons observed recently in heavy ion collisions at RHIC.

%In this note we discuss the stringy 2-form  current in QCD vacuum at $T=0$ 
%at weak magnetic field . The current 
%emerges from a polarization of the chiral condensate by the external magnetic field 
%and is proportional to the magnetic  susceptibility
%of the chiral condensate. The candidates for the light degrees of freedom 
%which could be the carriers of the stringy current are 
%considered. We suggest the interpretation of the QCD stringy current  as 2-form current in the
%particular limit of the recently formulated  1-form hydrodynamics  and relate the density 
%of strings  with the chiral condensate.

\end{abstract}
\maketitle

\section{Introduction}

Since the early work of 't Hooft \cite{thooft} and Mandelstam \cite{mandelstam}, it is widely believed that the QCD vacuum in the confined phase can be viewed as a dual superconductor. Due to the dual Meissner effect, the chromoelectric field gets repelled by the condensate of chromomagnetic 
degrees of freedom, and the emerging chromoelectric strings confine the quarks and bind them into mesons and baryons. While this qualitative picture is simple and attractive, it is not yet clear how to realize it microscopically in QCD. In particular, the nature of the chromomagnetic objects that condense in the vacuum is still not entirely understood, in spite of significant advances made over the last decades (see e.g. \cite{Shuryak:2018fjr,Chernodub:2008iv} and references therein). For example, apart from magnetic monopoles proposed originally, the confinement can also arise from the condensation of closed chromomagnetic strings  \cite{polyakov}.

In Londons' theory of conventional superconductivity, the flux of magnetic field inside an Abrikosov  vortex is surrounded by the electric supercurrent 
proportional to the condensate that screens the magnetic field in the bulk of a superconductor. In the dual superconductor model of confinement, it is thus natural to expect that the confining chromoelectric flux is surrounded by the supercurrent of chromomagnetic charges that shields the vacuum from the chromoelectric field. Does this supercurrent exist in QCD? If so, what are the consequences for the structure of hadrons? The answers to these questions are still lacking.

Quarks interact with the chromomagnetic degrees of freedom, and this interaction should affect both the chiral condensate and the hadrons. Since quarks, in addition to color charges, possess also the electric charge, they respond to an external magnetic field that can thus be used as a probe of non-perturbative QCD dynamics.

Indeed, an external Abelian magnetic field has emerged as a powerful probe of the QCD vacuum \cite{Kharzeev:2012ph,kharzeev2,Miransky:2015ava}. 
%There exists an evidence from the lattice studies \cite{lattice} that the QCD 
% vacuum is diamagnetic, and hence repels the applied magnetic field. 
 In chiral theory, one expects that the chiral condensate increases \cite{smilga1} in an external magnetic field (in accord with the ``magnetic catalysis" scenario \cite{Gusynin:1995nb,Miransky:2015ava}),  
and a constant density of magnetic moment gets generated \cite{smilga2}. At finite chemical potentials for the chiral and vector charges, the chiral magnetic effect \cite{cme} and chiral separation effect \cite{Son:2004tq} are induced, see \cite{Kharzeev:2012ph,Kharzeev:2013ffa} for reviews.
These effects have been discussed not only in the deconfined phase, but also 
in confined, chirally broken phases, see e.g. \cite{Kenji,Basar:2013qia,zakharov}. It was argued that 
in confined phase pionic effective strings could play an important
role by providing their core for the propagation of dissipationless chiral currents \cite{fukushima}.  
\vskip0.3cm

In this paper, we focus on a vacuum tensor current that emerges at zero temperature and zero chemical potential in an external abelian magnetic field. The non-vanishing v.e.v. of this tensor current gives rise to both the dipole magnetic moment of the vacuum and the current circulating in the plane perpendicular to the external magnetic field. The existence of this tensor supercurrent can be implied from the relations between the magnetic \cite{gkkv} and vortical \cite{afgk} susceptibilities of the quark condensate and the quantum anomalies. 
Once the term describing the anomalous response of the quark condensate is added to the effective chiral Lagrangian  \cite{gkkv}, it leads to the tensor current 
\beq\label{sc1}
J_{\mu\nu}=\chi <\bar{\Psi}\Psi>F_{\mu\nu},
\eeq
where an external magnetic field is described by the field strength tensor $F_{\mu\nu}$, and $<\bar{\Psi}\Psi>$ is the quark condensate. The value  of magnetic susceptibility of the condensate was discussed within several approaches; the Vainshtein relation \cite{Vainshtein:2002nv}
\beq\label{vain}
\chi = - \frac{N_c}{4 \pi^2 f_\pi^2} ,
\eeq
was derived from the VVA anomalous triangle diagram with the use of pion dominance in the axial channel.
\vskip0.3cm

In this paper we will address the following two questions:

\begin{enumerate}
\item What are the microscopic carriers of  the tensor current in confined QCD? Since the current can be excited by an arbitrarily weak magnetic field, it has to be carried by very light degrees
of freedom.
 
\item The tensor supercurrent is proportional to the chiral condensate
which breaks the global chiral symmetry. Can we interpret  this current in analogy with superfluidity or superconductivity?  
Does this current emerge at the boundary of a domain where the chiral symmetry breaking is different from the bulk pattern, similar to 
 supercurrents around strings and vortices in superfluidity and superconductivity? 
\end{enumerate}

These questions can be addressed both in Euclidean and Minkowski spaces. The
Euclidean QCD vacuum with a quark condensate, according to the Casher-Banks relation \cite{casher}, is characterized by a finite density 
of quasi-zero Dirac operator eigenmodes which are delocalized in 4d Euclidean space-time.
It was shown in \cite{buividovich} that magnetic susceptibility of the 
quark condensate is saturated by the zero modes of 4D Dirac operator; therefore
the key contribution to the tensor current should involve the defects supporting such zero modes. 
The simplest Euclidean defect supporting fermionic zero mode is the instanton.
The behavior of the fermion in the background of a single instanton 
in the external magnetic field has been considered in \cite{dunne}. It was found that a single instanton gives rise to a dipole electric moment of the quark quasi-zero modes, in agreement with the lattice QCD study \cite{buividovich2}; see \cite{Faccioli:2004ys} for a related observation for a polarized nucleon. 
Since an anti-instanton develops an electric dipole moment of an opposite orientation, it can be expected that a pair of an instanton and an anti-instanton develops a tensor electric moment, in accord with the emergence of the tensor current from the effective theory. While the microscopic picture can become quite complicated due to the instanton--anti-instanton interactions, the effective chiral Lagrangian allows to fix the magnitude of the tensor current in terms of the quark condensate. 
\vskip0.3cm

We advocate here the following interpretation: in an external magnetic field, the tensor supercurrent  in 4D Euclidean space-time screens the instanton--anti-instanton $I\bar{I}$ molecules in the vacuum ensemble. Indeed, the $I\bar{I}$ molecule supports the fermion zero modes \cite{shuryak} that in an external magnetic field develop the tensor electric moment \cite{dunne}. Since this tensor moment should be absent in the empty vacuum surrounding the molecule, it is screened by the tensor supercurrent.
In other words, one can say that magnetic field probes
the instanton molecule component of the QCD vacuum and induces the tensor supercurrent surrounding the individual molecules.

\vskip0.3cm

The paper is organized as follows. First we describe how the 2-form supercurrent 
emerges in confined phase of QCD within the low-energy effective theory in Section 2. In Section 3 we consider the microscopic 
aspects of tensor supercurrent and argue its relevance for the screening of the $I\bar{I}$
molecules. In Section 4 we describe the possible experimental manifestations of the tensor supercurrent. The comparison with the screening currents familiar in superconductivity and superfluidity, 
as well as interpretation of the tensor supercurrent as a conserved 2-form current of broken 1-form
global symmetry in hydrodynamics is presented in Discussion. Some open questions are formulated in Conclusion.

\section{Tensor supercurrent in the confined phase of QCD}

\subsection{Tensor currents and dipole moments}

In this Section we explain how the 2-form currents emerge
from the polarization of the chiral condensate \cite{gkkv,afgk}.
In hadronic phase of QCD, the chiral 
condensate $<\bar{\Psi}\Psi>$ breaks the $SU(N_f)_L\times SU(N_f)_R$ symmetry of the Lagrangian to the diagonal $SU(N_f)$ subgroup. 
Let us consider the response of this 
chiral condensate to an external electromagnetic field:
\beq
<0|\bar{\Psi}_f\sigma_{\mu\nu}\Psi_f|0> = \chi e_f<\bar{\Psi}\Psi>F_{\mu\nu}, 
\label{sus}
\eeq
and 
\beq
<0|\bar{\Psi}_f\sigma_{\mu\nu} \gamma_5\Psi_f|0> = \tilde{\chi}e_f <\bar{\Psi}\Psi>\tilde{F}_{\mu\nu} ,
\label{sus2}
\eeq
where $e_f$ is the electric charge of the quark with flavor $f$, and $\sigma_{\mu\nu}=\frac{1}{2i} [\gamma_\mu, \gamma_\nu]$  is the relativistic spin operator.
Since in four dimensions
\beq
\sigma_{\mu\nu}= i \epsilon_{\mu\nu\alpha\beta} \sigma_{\alpha\beta}\gamma_5 ,
\label{sigma}
\eeq
the electric and magnetic susceptibilities of the condensate $\chi$ and $\tilde{\chi}$  are related.
The  value of magnetic susceptibility introduced in \cite{smilga2} has been derived from the anomalous 
$<VVA>$ triangle
\cite{Vainshtein:2002nv} (as given by (\ref{vain})), in holography via 5d Chern-Simons term \cite{gk}, in an extended holographic
model \cite{afgk,harvey} and in lattice QCD \cite{buividovich}.

The physical interpretation of the vacuum tensor currents (\ref{sus}) and (\ref{sus2}) is as follows:
the $\bar{\Psi}_f\sigma_{0i}\Psi_f$ component of (\ref{sus}) that is a tensor charge corresponds to the electric dipole moment 
in the external electric field $E_i = F_{0i}$. Similarly, the $\bar{\Psi}_f\sigma_{0i}\gamma_5\Psi_f$ component of (\ref{sus2}) corresponds to the magnetic dipole
moment in the external magnetic field $B_i = \tilde{F_{oi}}$. Due to the kinematic relation (\ref{sigma}),  the non-vanishing dipole magnetic moment 
implies the ``electric" spatial tensor current $\sim \bar{\Psi}_f\sigma_{ji}\Psi_f$ in the plane transverse to the applied magnetic field $B_i$ 
while similarly the electric dipole moment implies the ``magnetic" spatial tensor current $\sim \bar{\Psi}_f\sigma_{ji} \gamma_5\Psi_f$ in the plane transverse to the applied electric field $E_i$.

Since there is no CP violation in QCD, there are no CP-odd terms in  (\ref{sus}) and (\ref{sus2}) -- for
instance, there is no induced electric dipole moment in a magnetic field. However the 
electric dipole moment squared does not vanish due to fluctuations, as was demonstrated in the lattice 
study \cite{buividovich2}. The fluctuations of the electric dipole moment emerge for example from an ensemble of instantons and anti-instantons that have opposite electric dipole moments in an external magnetic field \cite{dunne}. While the 
total electric dipole moment vanishes upon the averaging over the instanton--anti-instanton
ensemble, the correlator of the electric dipole moments does not vanish due
to the non-vanishing correlator of topological charges. The emergence of the spatial electric tensor current in the plane transverse to an external magnetic field in the presence of an instanton-antiinstanton pair is illustrated in Fig.1.

\begin{figure}[t]
\centering
\hspace{-3cm}
\begin{minipage}{.7\textwidth}
\includegraphics[scale=0.4]{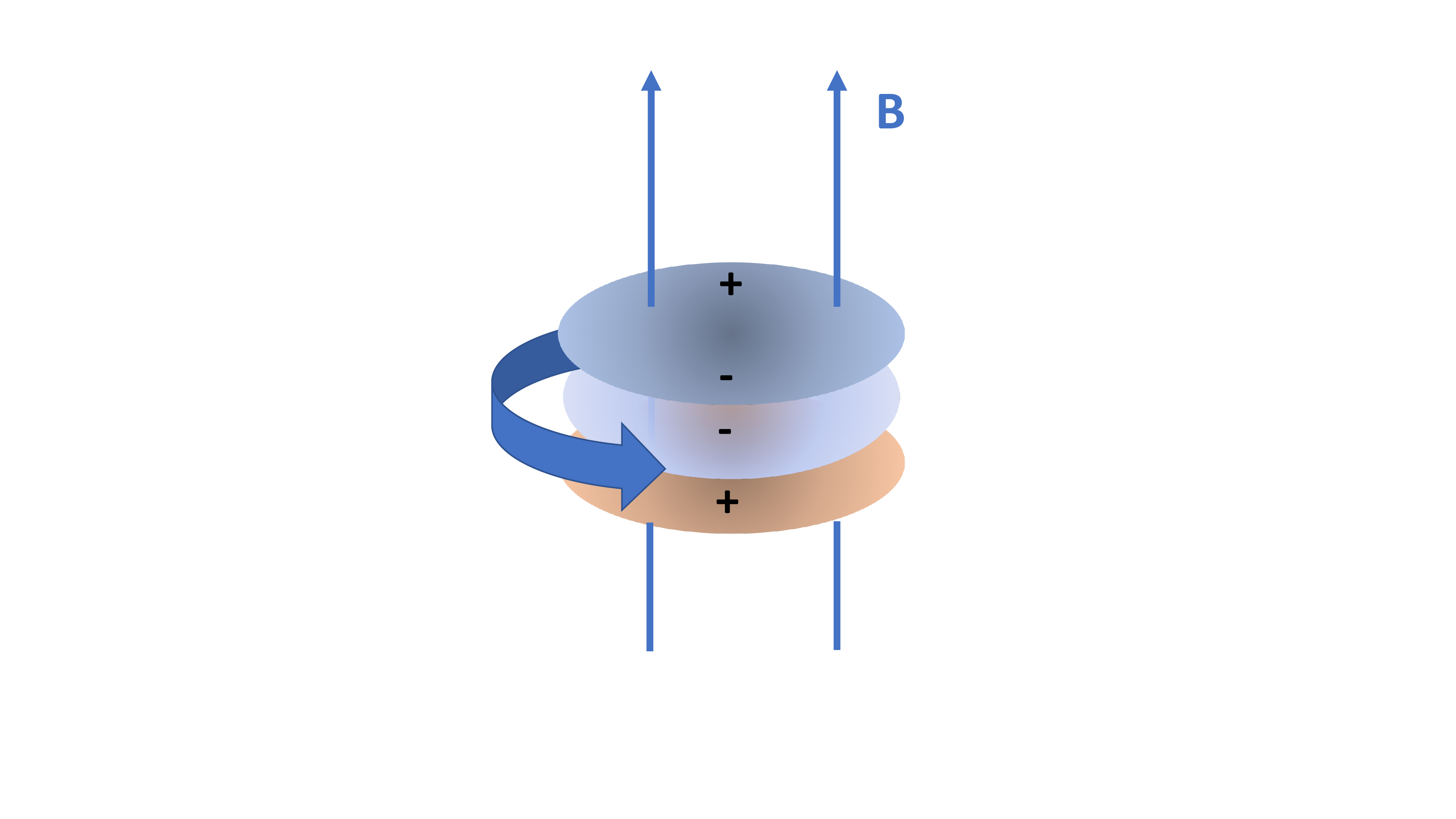}
\end{minipage}
\caption{Tensor supercurrent in the QCD vacuum. The external magnetic field induces electric dipole moments of opposite orientation in the instanton and anti-instanton, creating a tensor polarization of the quark zero modes. This tensor polarization is absent in the ``empty" vacuum surrounding the instanton-antiinstanton pair, and is thus screened by the tensor supercurrent proportional to the quark condensate.}
\label{fig1}
\end{figure}

\vskip0.3cm

The duality between electric and magnetic dipole moments is similar to the familiar duality between the vector and axial currents in the Schwinger model, the $(1+1)$-dimensional QED. In this model there exist
the vector and axial vacuum currents in the external gauge field that at one-loop level are given by 
\beq
<0|\bar{\Psi}_f\gamma_{\nu}\Psi_f|0> \sim A_{\nu} ,
\label{vector}
\eeq
\beq
<0|\bar{\Psi}_f\gamma_{\nu} \gamma_5\Psi_f|0> \sim \epsilon_{\mu \nu} A_{\mu},
\label{axial}
\eeq
which are related kinematically due to the identity
\beq
\gamma_{\nu} = i \epsilon_{\mu \nu}\gamma_{\mu} \gamma_5 .
\eeq
This means that a non-vanishing vector charge in the external field implies the 
non-vanishing axial current and vice versa. This is 
similar to our four-dimensional case, where the non-vanishing
dipole magnetic moment in an external magnetic field implies the spatial electric tensor current.

\subsection{Tensor current from magnetic susceptibility}\label{comp}

It is useful to introduce the antisymmetric rank two external field tensor $B_{\mu\nu}$ as a source for the
microscopic quark  tensor current (\ref{sus})
\beq
 <0|\bar{\Psi}\sigma_{\mu\nu}\Psi|0>= \frac{\delta L}{\delta B_{\mu \nu}}
\label{def}
\eeq
and similarly external field pseudotensor $\tilde{B}_{\mu \nu}$ as the source
for the pseudotensor quark current (\ref{sus2}). Implementing (\ref{def}) amounts to the
additional terms in
the effective  Chiral Lagrangian which can be derived  
at the quark level from the triangle diagram involving the vector, tensor and scalar
currents.  Taking into account  non-vanishing magnetic susceptibility, the resulting anomalous term
in the Chiral Lagrangian in the external $B_{\mu\nu}$ field is \cite{gkkv}
\beq
L_{anom}= \chi <\bar{\Psi}\Psi>F_{\mu\nu}B_{\mu \nu}{\rm Tr} B Q (U+U^{-1})+ 
\tilde{\chi} <\bar{\Psi}\Psi>\tilde{F}_{\mu \nu}\tilde{B}_{\mu \nu}{\rm Tr} B Q (U+U^{-1}) ;
\label{anom}
\eeq
B is the flavor matrix and Q is charge matrix, $U=\exp(\frac{i\pi^at^a}{f_{\pi}})$. The 
anomalous term yields in particular an effective mass of the pion in external fields in the chiral limit
\beq\label{pion_mass}
m^2_{\pi, eff} = \chi <\bar{\Psi}\Psi>F_{\mu\nu}B_{\mu \nu}f_{\pi}^{-2}
\eeq
that arises from the tensor polarization of the quark condensate.
\vskip0.3cm

The Chiral Lagrangian can be derived 
in the holographic framework from the 5d gauge theory
with $SU(N_f)\times SU(N_f)$ gauge group. The stringy currents in the holographic 
picture appear if we take into account the specific mixed CS-like term
in the 5d bulk Lagrangian for the extended hard-wall model \cite{karch,harvey,gkkv}
\beq
\delta S= \int d^5x \sqrt{-g}Tr (X^{+}F_LB + BF_RX)
\eeq
where X is scalar in bifundamental representation, B is self-dual antisymmetric
rank 2 field in the bifundamental representation and $F_{L,R}$ are the field strengths
for the left and right gauge groups. At the boundary of the holographic
5d space it yields the corresponding term (\ref{anom}) in the Chiral Lagrangian.

The  B field is sourced by stringy degrees of freedom due to the 
term $\int B_{\mu\nu} d\Sigma_{\mu\nu}$ in the string worldsheet action. Therefore we could introduce the stringy current
\beq
J_{\mu\nu} = \frac{\delta L}{\delta B_{\mu\nu}}
\eeq
From the anomalous term in the Chiral Lagrangian
we immediately get  the conserved stringy supercurrent in an external magnetic
field:
\beq
J_{\mu\nu}=\chi <\bar{\Psi}\Psi>F_{\mu\nu} .
\eeq

The mixed anomalous term also yields the  vector current proportional to the chiral condensate
if the external rank-two pseudotensor field has non-vanishing curvature 
\beq
J_{\nu}=\chi <\bar{\Psi}\Psi> \epsilon_{\nu \mu \alpha \beta} \partial_{\mu}\tilde{B}_{\alpha \beta}
\eeq
Since  this current is proportional to the condensate, it can be considered as analog of 
 vector supercurrent.  Similarly there is an axial current proportional
to the curvature of the tensor field and the chiral
condensate.

%\subsection{Stringy current and the tensor charge of nucleon}
%\subsection{Dual representation}

\section{Towards a microscopic picture}

\subsection{Dual Lagrangians and examples of  tensor currents}

Let us present two examples in which the tensor stringy currents in an external
magnetic field can be constructed explicitly.
The stringy currents are usually hidden in the original 
formulation of the theory but emerge clearly  in the dual formulation.
The first example  concerns the Polyakov's $(2+1)$-dimensional compact QED \cite{Polyakov:1976fu}. In that theory
there is a natural 2-form current
\beq
J_{\mu\nu}= \epsilon_{\mu\nu\alpha}\partial_{\alpha}\phi ,
\eeq
where $\phi$ is the pseudoscalar that is dual to the photon:
\beq
\epsilon_{\mu\nu\alpha}\partial_{\alpha}\phi = F_{\mu\nu} .
\eeq

This current
is conserved perturbatively, apart from the points where the vortices-monopoles are localized, 
and it counts the number of strings. The natural microscopic carrier
of  this 2-form current is the string of finite length with monopole and anti-monopole
at its ends. It is the string that provides confinement in the theory, and we thus see clearly the  two roles played by the tensor current. The monopole-antimonopole
pair at the ends of the string amounts to the magnetic dipole structure described by the tensor current. On the other hand, this tensor current is carried by the confining string -- this suggests a stringy interpretation for it.
\vskip0.3cm

The second example concerns the Abelian Higgs model in $(3+1)$-dimensional space-time. The Lagrangian
reads  
\beq
L= -\frac{1}{4}F^2 - |\partial_{\mu} -igA_{\mu}\phi|^2 - (g|\phi|^2 - v^2)^2 ,
\eeq
where the potential supports the v.e.v. of the scalar field. There are effective 
strings in this theory, and the v.e.v. of the scalar field vanishes at their cores. It is
useful to introduce the following parametrization for the complex scalar
\beq
\phi=\rho e^{i\theta} ;
\eeq
the scalar field $\phi$ can then be dualized into the 2-form gauge field $B_{\mu\nu}$, 
and the dual Lagrangian reads as 
\beq
L_{dual}= \frac{1}{2\rho^2} (dB)^2 -\frac{1}{4}F^2 - g\epsilon_{\mu\nu\alpha\beta}B_{\mu\nu}F_{\alpha\beta}
-\frac{1}{2} \epsilon_{\mu\nu\alpha\beta}B_{\mu\nu}\partial_{\alpha}\partial_{\beta} \theta ,
\eeq
where the last term corresponds to the interaction of the 2-form field with the effective strings.

Now we can define the stringy current in the standard manner in terms of this Lagrangian. If we consider the vacuum currents
we can put $\rho=v$ ($v$ is the v.e.v. of the $\rho$ field), and, to keep the analogy with QCD more close, rescale the 2-form field as 
\beq
\tilde{B}_{\mu\nu}= vB_{\mu\nu}
\eeq
The stringy dimension-3 pseudotensor current in the external magnetic field now reads as 
\beq
J_{\mu\nu} =gv\tilde{F}_{\mu\nu},
\eeq
which can be seen as the analog of the tensor current (\ref{sc1})  in low-energy
QCD that we discussed above. 

The microscopic interpretation of this current is
subtle. However one can view the scalar condensate in the vacuum in the Abelian Higgs model as a condensate of small closed loops of magnetic flux. In such a vacuum it is natural to assume 
that it is these small closed strings that get excited by magnetic field and 
provide a 2-form current.

\subsection{Euclidean picture. Stringy current and the Dirac operator spectrum.}

In Euclidean space, it is useful to interpret magnetic susceptibility 
in terms of the spectrum of 4D Euclidean Dirac operator \cite{buividovich} :
\beq
\hat{D}(A)\psi_n=i\lambda_n\psi_n
\eeq
which coincides with Dirac equation with an imaginary fermion 
mass $m=i\lambda$.
The spectral density is defined as 
\beq
\rho(\lambda)=<\sum_n \delta(\lambda- \lambda_n)>_{QCD} ;
\eeq
in the confined phase, 
according to the Casher-Banks relation, it is related to the quark condensate \cite{casher}:
\beq
<\bar{\Psi}\Psi>= \Sigma =\frac{\pi \rho(0)}{V}  .
\eeq

It was found in \cite{buividovich} that magnetic susceptibility 
of the condensate  has the following representation in terms of the Dirac operator spectrum 
\beq
<0|\bar{\Psi}_f\sigma_{\mu\nu}\Psi_f|0> = {\rm lim} _{\lambda\rightarrow 0} <\rho(\lambda)
\int d^4x \bar{\Psi}_{\lambda}\sigma_{\mu\nu}\Psi_{\lambda}>_{QCD} ,
\label{zero}
\eeq
which involves the tensor current of  zero  modes of the Dirac operator.
Assuming factorization (which has been checked numerically in \cite{buividovich}) on the r.h.s. of  (\ref{zero}),
we arrive at the conclusion that the tensor current is saturated 
by the zero modes of the Dirac operator in the external magnetic field 
in the background of $I\bar{I}$ molecule.  We have already advocated for this interpretation above, see also Fig.1. 
%Presumably the key role in the generation of the magnetic susceptibility
%is played by the t'Hooft instanton induced vertex involving fermionic zero modes.
%In (\ref{zero}) two fermionic legs from the t'Hooft vertex are involved
%in the condensate while two in the zero modes tensor current.

%Hence in the Euclidean setup the tensor current originates from topological defects
%supporting the zero modes of Dirac operator in magnetic field. There
%are at least two possible types of defects at weak magnetic field -- instantons 
%and magnetic strings. The discussion concerning the contribution 
%of zero modes on individual instantons into the tensor current
%can be found in \cite{dunne}. It was found there that indeed there 
%is some contribution linear in the external field. However in the
%QCD vacuum considered as the instanton-antiinstanton liquid there 
%are no individual zero modes at the instanton and all modes are 
%delocalized in the 4D Euclidean space-time. Moreover if we assume
%that the stringy current is supported by the instanton it is unclear
%what kind of screening of the chiral condensate takes place.

\subsection{Minkowski space. Light degrees of freedom?}

We have discussed above the nature of the stringy tensor current in Euclidean space-time.
However it is important to identify the carriers of
tensor current in the QCD ground state in Minkowski space. 
Because the tensor current appears as a response to even very weak magnetic field
$eB<<\Lambda_{QCD}^2$, the carriers of the current have to be very light. 
 On the other hand, the Euclidean
analysis  suggests the dominant role of some kind of topological objects, since the spectral Dirac operator representation of the tensor current is saturated by the quark zero modes.

Two possibilities that come to mind are:

{\it i)} The loop corrections to the ground state energy 
get deformed by the magnetic field and
provide the magnetic moment and the tensor current;

{\it ii)} There are specific stable light extended semiclassical configurations
which can be excited in a weak magnetic field and serve as the carriers
or the stringy tensor current. The example of rearrangement of the ground
state at very weak magnetic field has been discussed in \cite{ss} 
where the stabilization of the pionic domain walls has been demonstrated.

The scenario based on the virtual correction generalizes the evaluation
of the condensate in a magnetic field \cite{smilga1}.
In \cite{smilga1}, the loop of charged pions was found to generate the leading 
correction to the condensate in magnetic field. In our case, we have to analyze the
dependence of the vacuum energy on the external $B_{\mu\nu}$ field
through the one-loop correction to the  chiral lagrangian. The anomalous term in the
Lagrangian yields the effective pion mass (\ref{pion_mass}); therefore, one contribution of the desired type can be found by substituting this mass into the Heisenberg-Euler lagrangian in the 
external magnetic field.  

The contribution of the pion loop to the vacuum energy density in an external magnetic field is given by 
the Heisenberg-Euler theory as 
\beq
\epsilon_{vac}= - \frac{1}{16\pi^2} \int ^{\infty}_{0} \frac{d s}{s^3}\ \exp[-m_{\pi,eff}^2 s]\left[\frac{eBs}{\sinh (eBs)} -1\right] ,
\eeq
where the integral is over the Schwinger's proper time $s$. 
Substituting into this expression the effective mass of the pion in an external magnetic
field (\ref{pion_mass}) and taking derivative with respect to the external $B_{\mu\nu}$ field (not to be confused with magnetic field $B$),  
we get the tensor current up to the second power of the magnetic field:
\beq\label{tenpi}
J_{\mu\nu}= \chi <\bar{\Psi}\Psi>F_{\mu\nu} \left(1 +  \frac{e B \ln2}{16\pi^2 f_{\pi}^2}\right) .
\eeq

The physical interpretation of this formula is the following. The first term, which is already familiar to us,  describes the virtual charged pions that rotate in magnetic field on Landau levels and give rise to the magnetic moment of the vacuum.  The second term describes the correction that arises from the polarization of the pion, which is a composite particle of size $R_\pi \simeq (4\pi f_\pi)^{-1} \simeq 0.2$ fm, by the magnetic field. It is clear from (\ref{tenpi}) that it represents an expansion in powers of the ratio $R_\pi/R_B$ of the pion size to the magnetic length $R_B = (eB)^{-1/2}$. 

The quarks inside the pion are confined, so the second term in (\ref{tenpi}) can also be seen as originating from a tensor current carried by a string with quark and antiquark at its ends. It has been argued \cite{kharzeev} that the chiral condensate gets suppressed by the confining string - therefore it is not surprising to find a correction to the condensate response that originates from the composite nature of the pion.
\vskip0.3cm

In addition to the polarization of the charged pion, there exists also a contribution to the tensor current that is induced by the 
$\pi^0$-photon mixing in the magnetic field. Iterating this vertex, 
we will get the loop involving the photon
and pion propagators.   This opens an interesting possibility to describe the tensor current in terms of neutral pions. Indeed, it was shown in \cite{ss}
that low-energy QCD in an external magnetic field is unstable with respect to formation of the
$\pi^0$ domain walls. In the limit
of the massless quarks, the ground state is always unstable, 
the domain walls carry the baryon charge and are more energetically favorable than the ordinary 
nuclear matter. Moreover, the domain wall with the disc topology is 
enclosed with the $\pi$-mesonic string which completely screens 
the induced baryonic charge of the wall \cite{fukushima}. 

Very recently such pancake configurations were
suggested as the candidates for high spin baryons \cite{komar}.
The stabilization of the disc size occurs due to the edge chiral currents
at its boundary. We speculate that the polarization of the edge chiral currents at the boundary by magnetic field can also give rise to the tensor current, similarly to the case of charged pions considered above. It will be interesting to investigate this scenario further.

\section{Experimental manifestations}

\subsection{Vector meson spin alignment in heavy ion collisions}

Heavy ion collisions produce QCD matter subjected to strong magnetic field and large vorticity. 
In both cases, we expect the emergence of the tensor supercurrent. One manifestation of this tensor current is the polarization of $\Lambda$ hyperons that was estimated in \cite{afgk} for the case of finite vorticity. It was found that the resulting polarization is consistent with the experimental results of STAR Collaboration at RHIC. 

There exists another, potentially even more prominent, signature of the tensor supercurrent in heavy ion collisions. Indeed, consider a state $|\Omega>$ of QCD matter with a non-zero tensor charge (\ref{sus2}), i.e. the state with a finite magnetic dipole moment $<\Omega|\bar{\Psi}_f\sigma_{0i} \gamma_5\Psi_f|\Omega>$. The quantum numbers of this state suggest that it should hadronize with a copious emission of spin-polarized axial vector $J^{PC}=1^{+-}$ mesons $V$, such as $a_1(1260)$, $h_1(1170)$, $b_1(1235)$ and so on; the corresponding meson emission amplitude is
\beq
<\Omega|\bar{\Psi}_f\sigma_{\mu\nu} \gamma_5\Psi_f|A> = i f_A \left(\epsilon_\mu k_\nu - \epsilon_\nu k_\mu \right) ,
\eeq
where $f_A$ is the meson decay constant, $\epsilon_\mu$ is the meson polarization vector, and $k_\nu$ is the meson's four-momentum. 

Let us consider this amplitude in the rest frame of matter characterized by a dipole magnetic moment along the direction of the magnetic field (axis $i$). Assuming that the produced meson in this reference frame is slow, with three-momentum much smaller than the meson mass $M_A$, $k_i \ll M_A$, we find 
\beq
<\Omega|\bar{\Psi}_f\sigma_{0i} \gamma_5\Psi_f|A> \simeq - i f_A M_A\ \epsilon_i ,
\eeq
i.e. the axial-vector mesons are produced with polarization along the direction of the magnetic field. It may be hard to measure the polarization of the axial-vector mesons directly in heavy ion experiments, where only the polarization of vector mesons has been measured so far. Fortunately, for the mesons listed above the dominant decay mode is $A \to V + \pi$, i.e. a vector meson and a pion (or kaon, for strange axial-vector mesons). Since this is an $s$-wave decay, and pion has a zero spin, the produced vector mesons should inherit the polarization of the parent axial-vector mesons, and should thus be characterized by a spin alignment along the magnetic field. A remarkably strong spin alignment of vector mesons has been observed recently by STAR Collaboration at RHIC. It significantly exceeds the predictions based on both the statistical hadron model and the recombination of polarized quarks.

It is hard to estimate the vector meson polarization resulting from the tensor supercurrent in a model-independent way, but it should greatly exceed the predictions of statistical model where the polarization originates from the coupling of vector meson's spin to vorticity. Indeed, in the statistical model the vector mesons are excited states of the system, whereas the tensor current and the resulting polarization characterize its ground state, even at zero temperature.

\subsection{Tensor charge of the nucleon and transversity}

%\subsection{Transversity quark distributions in polarized deep-inelastic scattering}

The forward matrix element of the tensor current operator ${\hat J}_{\mu\nu} = \bar{\Psi}_f\sigma_{\mu\nu} \gamma_5\Psi_f$ over the transversely polarized nucleon defines the tensor charge of the nucleon:
\beq
< S^T|\bar{\Psi}_f\sigma_{\mu\nu} \gamma_5\Psi_f|S^T> = 2 \delta q_f \left(P_\mu S^T_\nu - P_\nu S^T_\mu \right),
\eeq
where $P_\mu$ is the four-momentum of the nucleon, $S^T$ describes the nucleon's transverse polarization, and $\delta q_f$ is quark transversity given by the integral over the difference of quark $\delta q_f (x)$ and antiquark $\delta {\bar q}_f (x)$ transversity distribution functions:
\beq
\delta q_f = \int_0^1 dx (\delta q_f(x) - \delta {\bar q}_f(x)), 
\eeq
where $x$ is Bjorken $x$.

In this case the role of magnetic field or vorticity is played by the spin of the nucleon. Nevertheless, 
just as the magnetic dipole moment of the vacuum is greatly enhanced by non-perturbative effects (tensor polarization of the quark condensate), we expect that a similar enhancement should occur for the tensor charge of the nucleon. Because this enhancement results microscopically from extended topological objects, transversity has to be a higher twist effect, in accord with perturbative analysis. Nevertheless, at moderate momentum transfers $Q^2 \sim 4 \pi f_\pi^2 \simeq 1$ GeV, basing on the analysis in Section \ref{comp}, we expect that transversity should be very large -- in accord with experimental observations. 

Moreover, in analogy with our discussion of the axial vector meson production in the decay of QCD matter, one can expect a strong coupling of axial vector mesons to the nucleons. The value of the corresponding coupling constant has been extracted from the data for the $a_1$ meson and is indeed large \cite{Gamberg:2001xc}:
\beq
g_{a_1NN} = 9.3 \pm 1 .
\eeq
We can compare this to the vector $\rho$ meson coupling to the nucleon, which is known to be significantly smaller \cite{Stoks:1996yj}:
\beq
g_{\rho NN} = 2.52 \pm 0.06 .
\eeq
\vskip0.3cm
To summarize this section, it appears that transversity and the tensor charge of the nucleon are enhanced by the quantum chiral anomaly.

\section{Discussion and future directions}

\subsection{Role of instanton molecules}

In a conventional superconductor, the persistent electric current can be expressed
in terms of the phase of charged condensate 
\beq
J_{\mu}= n_s (\partial_{\mu}\Phi - A_{\mu})
\eeq
where $n_s$ is proportional to the condensate of the Cooper pairs $<\Psi\Psi>$  and $\Phi$
is the phase of the order parameter. The charge current turns to be 
proportional to the gauge field in the London limit, the photon gets massive 
and magnetic field can not penetrate the superconductor due to the screening 
by electric supercurrent. The supercurrent  also screens the magnetic field inside
the Abrikosov vortex where the condensate vanishes in its core. The supercurrent
arises due to the abelian gauge symmetry in this case. The example of the supercurrent
arising from the global symmetry is provided by the superfluid where the superfluid velocity 
is proportional to the neutral condensate. The superfluid supercurrent
is proportional to the superfluid density and the gradient of the neutral condensate phase $\phi$:
\beq\label{sfluid}
J_{\nu}\propto \rho_{s}\ \partial_{\nu}\phi .
\eeq
This supercurrent screens the angular velocity (which can be treated as an external gravimagnetic field) resulting in the emergence of vortices.
Note that the superfluid density $\rho_s$ is
the analog of the topological susceptibility of the QCD vacuum.
\vskip0.3cm

The chiral symmetry is broken locally inside the instanton
molecule \cite{shuryak}; therefore necessarily there are gradients of the pion field which is the
phase of the chiral condensate at the boundary of the 4D region occupied by the 
molecule. The pions are pseudoscalar Goldstone bosons related to the spontaneous breaking of the
global chiral symmetry. The tensor supercurrent we are interested in would arise 
if there were vector (photon-like) Goldstone bosons associated with a broken 1-form symmetry. The possible mechanism for the emergence of these vector excitations is provided by the  $\pi^0$-photon mixing in the external magnetic field
due to the chiral anomaly. Hence the pseudoscalar and vector Goldstone modes are coupled in a magnetic field.
Within this interpretation our tensor current acquires the ``standard" interpretation (\ref{sfluid}), 
becoming a 2-form current of the vector Goldstone mode.

\vskip0.3cm

%We have two phenomena indicating the rearrangement of the ground state at arbitrary small
%magnetic field in the chiral limit. 
The tensor supercurrent (which screens the instanton
molecules in the Euclidean version) amounts to the alignment of the instanton molecules 
in the external magnetic field. A potentially related phenomenon in the language of the chiral effective theory is the emergence of a pancake-like 
configuration of pionic domain walls and anti-walls in Minkowski space in weak magnetic field \cite{ss}.
%Since there are a kind of "`wall-anti-wall"' structure with local restoration of chiral symmetry
%in the instanton molecule we could speculate that these two phenomena could be related.
Another related phenomenon discussed previously within the chiral effective theory is the anomaly-induced quadrupole moment of the neutron in an external magnetic field \cite{Kharzeev:2011sq}.

\subsection{Tensor supercurrent and magneto-hydrodynamics}

Recently  hydrodynamics involving electromagnetism  has been reformulated in an interesting way \cite{hofman,jain,son}
which is based on the 1-form global symmetry \cite{gaiotto} and the corresponding 
conserved 2-form stringy current. The 1-form symmetry is responsible for the conservation
of the number of strings and the Bianchi identity yields the conservation of 2-form current 
(in the absence of magnetic monopoles).  Two-form charge density is identified as  the density of 
magnetic  dipole charge which counts the 
number of magnetic field lines through an arbitrary surface. 

The fate of the global 1-form symmetry determines three different regimes of the theory:
\begin{enumerate}

\item The 1-form symmetry is unbroken;

\item The 1-form symmetry is partially broken;

\item The 1-form symmetry is completely broken.

\end{enumerate}
In these different symmetry breaking patterns one can obtain, respectively,  magneto-hydrodynamics (MHD), stringy
fluid, and bound-charge plasma (see \cite{jain} for the classification
of regimes and the connections between them). The vector Goldstone for the 1-form symmetry was 
identified with the photon.

Which situation gets realized in QCD? The quark tensor current in general is not conserved hence in the
quark sector the symmetry is broken. However it is conserved in the vacuum sector 
in the external magnetic field,  since we do not 
assume the presence of  magnetic monopoles -- indeed, the current (\ref{sc1}) is conserved due to the Bianchi identity. Hence we have the main
ingredient of the 1-form hydrodynamics in the vacuum  sector and may hypothesize that the QCD vacuum 
in magnetic field can be described as a kind of stringy superfluid. Within this regime of 1-form hydrodynamics 
the magnetic susceptibility of the condensate defines  the density of the instanton molecules. We postpone
discussion of the 1-form symmetry in the hydrodynamic approach to low-energy QCD in magnetic field for a separate study.

\section{Conclusion}

In this paper we have addressed the microscopic
origin of the  tensor supercurrent that arises in low-energy QCD in an external magnetic field or in a rotating frame and that is proportional to the
chiral condensate. We have 
suggested the following microscopic picture. There is a finite density 
of instanton--anti-instanton $I\bar{I}$ molecules in the QCD vacuum. As it was argued long time ago  \cite{shuryak}, 
%the chiral symmetry is restored completely or partially locally inside
%the molecule like restoration of symmetry occurs  inside the core of the 
%defect in superconductivity or superfluidity.
the instanton and anti-instanton host the pair of fermionic zero-modes. When
the external magnetic field is switched on, the fermion zero modes become polarized and develop a quadrupole moment that gets screened by the 
tensor supercurrent. All of the vacuum $I\bar{I}$ molecules become aligned in the
external magnetic field. Hence the external magnetic field probes
the molecule component of the QCD vacuum via the tensor supercurrent.

We have also argued that the tensor supercurrent manifests itself in experiment through the polarization of axial-vector and vector mesons in heavy ion collisions, and through the tensor charge of the nucleon and transversity in deep-inelastic scattering. All of these phenomena are induced by the chiral anomaly. The tensor supercurrent may play an important role 
in the stringy interpretation of magnetic hydrodynamics; we will discuss this interesting issue elsewhere.

The localization of zero modes of Dirac operator 
on extended defects has been 
investigated in lattice QCD. There were evidences that the Euclidean 4-dimensional 
zero modes ``live" on 2d surfaces and 3d volumes \cite{Horvath:2003yj,poli}
corresponding to the worldsheets of effective strings and effective
domain walls, correspondingly. In this study we considered the tensor supercurrents
screening the instanton molecules but similar
tensor current could exist on a pair of extended objects in the Euclidean space 
with the opposite topological charges.  

\vskip0.3cm

The stringy current considered here also has a counterpart in condensed matter physics. 
The chiral condensate in QCD is an analog of the neutral exciton condensate 
and the current we have considered corresponds to the polarization of the exciton condensate in magnetic field. Such currents have been indeed 
discussed \cite{macdon,lozovik}, so it would be interesting to pursue this 
analogy further. One more open question concerns the possible role of the charged tensor 
current discussed in the holographic framework in \cite{Gorsky:2017sgy}.

\vskip0.3cm

We  thank  A. Cherman, Z. Komargodski and E. Shuryak for
useful discussions. The work of A.G. was supported
by Basis Foundation fellowship and RFBR grant 19-02-00214. 
A.G. thanks Simons Center for Geometry and Physics at Stony Brook
University, Kavli Institute at UCSB and FTPI at University of Minnesota
where parts of the work have been carried out for the hospitality and support.
The work of D.K. was supported by the U.S. Department of Energy, Office of Nuclear Physics, under contracts DE-FG-88ER40388 and DE-AC02-98CH10886


\begin{thebibliography}{99}
\bibitem{thooft}
G. 't Hooft, in High Energy Physics, Ed. A. Zichichi, Editrice Compositori, Bolognia, (1976).
\bibitem{mandelstam}
S. Mandelstam, %"`Vortices And Quark Confinement In Nonabelian Gauge Theories"'
 Phys. Lett, B53, 476 (1975).
 %\cite{Shuryak:2018fjr}
\bibitem{Shuryak:2018fjr} 
  E.~Shuryak,
  %``Lectures on nonperturbative QCD,''
  arXiv:1812.01509 [hep-ph];\\
  J.~Liao and E.~Shuryak,
  %``Strongly coupled plasma with electric and magnetic charges,''
  Phys.\ Rev.\ C {\bf 75}, 054907 (2007)
  doi:10.1103/PhysRevC.75.054907
  [hep-ph/0611131].
  %%CITATION = doi:10.1103/PhysRevC.75.054907;%%
  %159 citations counted in INSPIRE as of 21 Jan 2020
  %%CITATION = ARXIV:1812.01509;%%
  %1 citations counted in INSPIRE as of 20 Jan 2020
  %\cite{Chernodub:2008iv}
\bibitem{Chernodub:2008iv} 
  M.~N.~Chernodub, A.~Nakamura and V.~I.~Zakharov,
  %``Manifestations of magnetic vortices in equation of state of Yang-Mills plasma,''
  Phys.\ Rev.\ D {\bf 78}, 074021 (2008)
  doi:10.1103/PhysRevD.78.074021
  [arXiv:0807.5012 [hep-lat]].
  %%CITATION = doi:10.1103/PhysRevD.78.074021;%%
  %26 citations counted in INSPIRE as of 20 Jan 2020
 
 
\bibitem{polyakov} 
  A.~M.~Polyakov,
 % ``Confining strings,''
  Nucl.\ Phys.\ B {\bf 486}, 23 (1997)
  doi:10.1016/S0550-3213(96)00601-3
  [hep-th/9607049].
  
  %\cite{Kharzeev:2012ph}
\bibitem{Kharzeev:2012ph} 
  D.~E.~Kharzeev, K.~Landsteiner, A.~Schmitt and H.~U.~Yee,
  %``'Strongly interacting matter in magnetic fields': an overview,''
  Lect.\ Notes Phys.\  {\bf 871}, 1 (2013)
  doi:10.1007/978-3-642-37305-3-1
  [arXiv:1211.6245 [hep-ph]].
  %%CITATION = doi:10.1007/978-3-642-37305-3_1;%%
  %222 citations counted in INSPIRE as of 20 Jan 2020
  
  \bibitem{kharzeev2} 
  D.~E.~Kharzeev,
 % ``Topology, magnetic field, and strongly interacting matter,''
  Ann.\ Rev.\ Nucl.\ Part.\ Sci.\  {\bf 65}, 193 (2015)
  doi:10.1146/annurev-nucl-102313-025420
  [arXiv:1501.01336 [hep-ph]].	
  
  %\cite{Miransky:2015ava}
\bibitem{Miransky:2015ava} 
  V.~A.~Miransky and I.~A.~Shovkovy,
  %``Quantum field theory in a magnetic field: From quantum chromodynamics to graphene and Dirac semimetals,''
  Phys.\ Rept.\  {\bf 576}, 1 (2015)\\
  doi:10.1016/j.physrep.2015.02.003
  [arXiv:1503.00732 [hep-ph]].
  %%CITATION = doi:10.1016/j.physrep.2015.02.003;%%
  %297 citations counted in INSPIRE as of 20 Jan 2020

  
%\bibitem{lattice}
%Diamagnetic vacuum
\bibitem{smilga1} 
  I.~A.~Shushpanov and A.~V.~Smilga,
 % ``Quark condensate in a magnetic field,''
  Phys.\ Lett.\ B {\bf 402}, 351 (1997)
  doi:10.1016/S0370-2693(97)00441-3
  [hep-ph/9703201].

%\cite{Gusynin:1995nb}
\bibitem{Gusynin:1995nb} 
  V.~P.~Gusynin, V.~A.~Miransky and I.~A.~Shovkovy,
  %``Dimensional reduction and catalysis of dynamical symmetry breaking by a magnetic field,''
  Nucl.\ Phys.\ B {\bf 462}, 249 (1996)
  doi:10.1016/0550-3213(96)00021-1
  [hep-ph/9509320].
  %%CITATION = doi:10.1016/0550-3213(96)00021-1;%%
  %423 citations counted in INSPIRE as of 20 Jan 2020

\bibitem{smilga2} 
  B.~L.~Ioffe and A.~V.~Smilga,
%  ``Nucleon Magnetic Moments and Magnetic Properties of Vacuum in QCD,''
  Nucl.\ Phys.\ B {\bf 232}, 109 (1984).
  doi:10.1016/0550-3213(84)90364-X

\bibitem{cme}
	 K.~Fukushima, D.~E.~Kharzeev and H.~J.~Warringa,
%  ``The Chiral Magnetic Effect,''
  Phys.\ Rev.\ D {\bf 78}, 074033 (2008)
  doi:10.1103/PhysRevD.78.074033
  [arXiv:0808.3382 [hep-ph]].

%\cite{Son:2004tq}
\bibitem{Son:2004tq} 
  D.~T.~Son and A.~R.~Zhitnitsky,
  %``Quantum anomalies in dense matter,''
  Phys.\ Rev.\ D {\bf 70}, 074018 (2004)\\
  doi:10.1103/PhysRevD.70.074018
  [hep-ph/0405216].
  %%CITATION = doi:10.1103/PhysRevD.70.074018;%%
  %245 citations counted in INSPIRE as of 20 Jan 2020

%\cite{Kharzeev:2013ffa}
\bibitem{Kharzeev:2013ffa} 
  D.~E.~Kharzeev,
  %``The Chiral Magnetic Effect and Anomaly-Induced Transport,''
  Prog.\ Part.\ Nucl.\ Phys.\  {\bf 75}, 133 (2014)\\
  doi:10.1016/j.ppnp.2014.01.002
  [arXiv:1312.3348 [hep-ph]].
  %%CITATION = doi:10.1016/j.ppnp.2014.01.002;%%
  %237 citations counted in INSPIRE as of 20 Jan 2020	
  
	
	\bibitem{Kenji}
	K.~Fukushima and K.~Mameda,
%  ``Wess-Zumino-Witten action and photons from the Chiral Magnetic Effect,''
  Phys.\ Rev.\ D {\bf 86}, 071501 (2012)
  doi:10.1103/PhysRevD.86.071501
  [arXiv:1206.3128 [hep-ph]].
  
  %\cite{Basar:2013qia}
\bibitem{Basar:2013qia} 
  G.~Basar, D.~E.~Kharzeev and I.~Zahed,
  %``Chiral and Gravitational Anomalies on Fermi Surfaces,''
  Phys.\ Rev.\ Lett.\  {\bf 111}, 161601 (2013)\\
  doi:10.1103/PhysRevLett.111.161601
  [arXiv:1307.2234 [hep-th]].
  %%CITATION = doi:10.1103/PhysRevLett.111.161601;%%
  %38 citations counted in INSPIRE as of 20 Jan 2020
  
	\bibitem{zakharov}
	A.~Avdoshkin, A.~V.~Sadofyev and V.~I.~Zakharov,
 % ``IR properties of chiral effects in pionic matter,''
  Phys.\ Rev.\ D {\bf 97}, no. 8, 085020 (2018)
  doi:10.1103/PhysRevD.97.085020
  [arXiv:1712.01256 [hep-ph]].
\bibitem{fukushima} 
  K.~Fukushima and S.~Imaki,
 % ``Anomaly inflow on QCD axial domain-walls and vortices,''
  Phys.\ Rev.\ D {\bf 97}, no. 11, 114003 (2018)\\
  doi:10.1103/PhysRevD.97.114003
  [arXiv:1802.08096 [hep-ph]].




\bibitem{gkkv} 
  A.~Gorsky, P.~N.~Kopnin, A.~Krikun and A.~Vainshtein,
%  ``More on the Tensor Response of the QCD Vacuum to an External Magnetic Field,''
  Phys.\ Rev.\ D {\bf 85}, 086006 (2012)
  doi:10.1103/PhysRevD.85.086006
  [arXiv:1201.2039 [hep-ph]].
  
\bibitem{afgk} 
  A.~Aristova, D.~Frenklakh, A.~Gorsky and D.~Kharzeev,
%  ``Vortical susceptibility of finite-density QCD matter,''
  JHEP {\bf 1610}, 029 (2016)
  doi:10.1007/JHEP10(2016)029
  [arXiv:1606.05882 [hep-ph]].	
  
  %\cite{Vainshtein:2002nv}
\bibitem{Vainshtein:2002nv} 
  A.~Vainshtein,
  %``Perturbative and nonperturbative renormalization of anomalous quark triangles,''
  Phys.\ Lett.\ B {\bf 569}, 187 (2003)
  doi:10.1016/j.physletb.2003.07.038
  [hep-ph/0212231].
  %%CITATION = doi:10.1016/j.physletb.2003.07.038;%%
  %102 citations counted in INSPIRE as of 20 Jan 2020
  
	\bibitem{casher} 
  T.~Banks and A.~Casher,
% ``Chiral Symmetry Breaking in Confining Theories,''
  Nucl.\ Phys.\ B {\bf 169}, 103 (1980).
  doi:10.1016/0550-3213(80)90255-2
\bibitem{shuryak}
T. Schaefer, E. V. Shuryak, J. J. M. Verbaarschot 
%"`The Chiral phase transition and instanton - anti-instanton molecules"'
Phys. Rev. D51, 1267 (1995)

	
	
\bibitem{buividovich} 
  P.~V.~Buividovich, M.~N.~Chernodub, E.~V.~Luschevskaya and M.~I.~Polikarpov,
 % ``Chiral magnetization of non-Abelian vacuum: A Lattice study,''
  Nucl.\ Phys.\ B {\bf 826}, 313 (2010)
  doi:10.1016/j.nuclphysb.2009.10.008
  [arXiv:0906.0488 [hep-lat]].
  
\bibitem{dunne} 
  G.~Basar, G.~V.~Dunne and D.~E.~Kharzeev,
 % ``Electric dipole moment induced by a QCD instanton in an external magnetic field,''
  Phys.\ Rev.\ D {\bf 85}, 045026 (2012)
  doi:10.1103/PhysRevD.85.045026
  [arXiv:1112.0532 [hep-th]].
	\bibitem{buividovich2}
	 P. V. Buividovich, M. N. Chernodub, E. V. Luschevskaya, M. I. Polikarpov, “Quark electric dipole moment induced by
magnetic field,” Phys. Rev.D81
, 036007 (2010). 

%\cite{Faccioli:2004ys}
\bibitem{Faccioli:2004ys} 
  P.~Faccioli,
  %``Strong CP breaking and quark antiquark repulsion in QCD, at finite theta,''
  Phys.\ Rev.\ D {\bf 71}, 091502 (2005)
  doi:10.1103/PhysRevD.71.091502
  [hep-ph/0404137].
  %%CITATION = doi:10.1103/PhysRevD.71.091502;%%
  %8 citations counted in INSPIRE as of 20 Jan 2020
  
  %\cite{Polyakov:1976fu}
\bibitem{Polyakov:1976fu} 
  A.~M.~Polyakov,
  %``Quark Confinement and Topology of Gauge Groups,''
  Nucl.\ Phys.\ B {\bf 120}, 429 (1977).
  doi:10.1016/0550-3213(77)90086-4
  %%CITATION = doi:10.1016/0550-3213(77)90086-4;%%
  %1542 citations counted in INSPIRE as of 20 Jan 2020
  
  \bibitem{ss}
  D.~T.~Son and M.~A.~Stephanov,
%  ``Axial anomaly and magnetism of nuclear and quark matter,''
  Phys.\ Rev.\ D {\bf 77}, 014021 (2008)
  [arXiv:0710.1084 [hep-ph]].
  
  %\cite{Kharzeev:2011sq}
\bibitem{Kharzeev:2011sq} 
  D.~E.~Kharzeev, H.~U.~Yee and I.~Zahed,
  %``Anomaly-induced Quadrupole Moment of the Neutron in Magnetic Field,''
  Phys.\ Rev.\ D {\bf 84}, 037503 (2011)
  doi:10.1103/PhysRevD.84.037503
  [arXiv:1104.0998 [hep-ph]].
  %%CITATION = doi:10.1103/PhysRevD.84.037503;%%
  %9 citations counted in INSPIRE as of 21 Jan 2020
  

	\bibitem{harvey} 
  S.~K.~Domokos, J.~A.~Harvey and A.~B.~Royston,
 % ``Successes and Failures of a More Comprehensive Hard Wall AdS/QCD,''
  JHEP {\bf 1304}, 104 (2013)
  doi:10.1007/JHEP04(2013)104
  [arXiv:1210.6351 [hep-th]].

\bibitem{gk} 
  A.~Gorsky and A.~Krikun,
%  ``Magnetic susceptibility of the quark condensate via holography,''
  Phys.\ Rev.\ D {\bf 79}, 086015 (2009)
  doi:10.1103/PhysRevD.79.086015
  [arXiv:0902.1832 [hep-ph]].
	
\bibitem{Gamberg:2001xc} 
  L.~P.~Gamberg and G.~R.~Goldstein,
%  ``Estimates of the nucleon tensor charge,''
  hep-ph/0106178.
  %%CITATION = HEP-PH/0106178;%%
  %10 citations counted in INSPIRE as of 01 Jan 2020	
  
  %\cite{Stoks:1996yj}
\bibitem{Stoks:1996yj} 
  V.~G.~J.~Stoks and T.~A.~Rijken,
 % ``Meson - baryon coupling constants from a chiral invariant SU(3) Lagrangian and application to N N scattering,''
  Nucl.\ Phys.\ A {\bf 613}, 311 (1997)
  doi:10.1016/S0375-9474(96)00462-9
  [nucl-th/9611002].


\bibitem{hofman}
S.~Grozdanov, D.~M.~Hofman and N.~Iqbal,
  %``Generalized global symmetries and dissipative magnetohydrodynamics,''
  Phys.\ Rev.\ D {\bf 95}, no. 9, 096003 (2017)
  doi:10.1103/PhysRevD.95.096003
  [arXiv:1610.07392 [hep-th]].
  %%CITATION = doi:10.1103/PhysRevD.95.096003;%%
  %58 citations counted in INSPIRE as of 21 Jan 2020
\\
D.~M.~Hofman and N.~Iqbal,
  %``Goldstone modes and photonization for higher form symmetries,''
  SciPost Phys.\  {\bf 6}, no. 1, 006 (2019)
  doi:10.21468/SciPostPhys.6.1.006
  [arXiv:1802.09512 [hep-th]].
  %%CITATION = doi:10.21468/SciPostPhys.6.1.006;%%
  %14 citations counted in INSPIRE as of 21 Jan 2020


\bibitem{son} 
  P.~Glorioso and D.~T.~Son,
%  ``Effective field theory of magnetohydrodynamics from generalized global symmetries,''
  arXiv:1811.04879 [hep-th].
	\bibitem{jain} 
	J.~Armas and A.~Jain,
  %``One-form superfluids & magnetohydrodynamics,''
  JHEP {\bf 2001}, 041 (2020)
  doi:10.1007/JHEP01(2020)041
  [arXiv:1811.04913 [hep-th]].
  %%CITATION = doi:10.1007/JHEP01(2020)041;%%
  %10 citations counted in INSPIRE as of 21 Jan 2020
  
\bibitem{gaiotto}
	D. Gaiotto, A. Kapustin, N. Seiberg and B. Willett,
%"`Generalized Global Symmetries"'
JHEP 02 (2015) 172.
	

\bibitem{komar}
	 Z.~Komargodski,
 % ``Baryons as Quantum Hall Droplets,''
  arXiv:1812.09253 [hep-th].
	
\bibitem{karch} 
  R.~Alvares, C.~Hoyos and A.~Karch,
%  ``An improved model of vector mesons in holographic QCD,''
  Phys.\ Rev.\ D {\bf 84}, 095020 (2011)
  doi:10.1103/PhysRevD.84.095020
  [arXiv:1108.1191 [hep-ph]].
	
	

  
	\bibitem{kharzeev} 
  D.~E.~Kharzeev and F.~Loshaj,
%  ``Partial restoration of chiral symmetry in a confining string,''
  Phys.\ Rev.\ D {\bf 90}, no. 3, 037501 (2014)
  doi:10.1103/PhysRevD.90.037501
  [arXiv:1404.7746 [hep-ph]].
	
%\cite{Horvath:2003yj}
\bibitem{Horvath:2003yj} 
  I.~Horvath, S.~J.~Dong, T.~Draper, F.~X.~Lee, K.~F.~Liu, N.~Mathur, H.~B.~Thacker and J.~B.~Zhang,
  %``Low dimensional long range topological charge structure in the QCD vacuum,''
  Phys.\ Rev.\ D {\bf 68}, 114505 (2003)
  doi:10.1103/PhysRevD.68.114505
  [hep-lat/0302009].
  %%CITATION = doi:10.1103/PhysRevD.68.114505;%%
  %116 citations counted in INSPIRE as of 21 Jan 2020

\bibitem{poli}
	J. Greensite, S. Olejnik, M.I. Polikarpov, S.N. Syritsyn and V.I. Zakharov,
	%"'Localized eigenmodes of covariant laplacians in the Yang-Mills vacuum"',
	Phys. Rev.D71(2005) 114507, [hep-lat/0504008].\\
	 V.~I.~Zakharov,
%  ``Dual string from lattice Yang-Mills theory,''
  AIP Conf.\ Proc.\  {\bf 756}, no. 1, 182 (2005)
  doi:10.1063/1.1920945
  [hep-ph/0501011].
  
	\bibitem{macdon}
%	"`How to make a bilayer exciton condensate flow"'
Jung-Jung Su, A.H. MacDonald 
	Nature Physics 4, 799 - 802 (2008),
	arXiv:0801.3694 [cond-mat.mes-hall]\\
%	"`Room-temperature superfluidity in graphene bilayers"'
Hongki Min, Rafi Bistritzer, Jung-Jung Su, and A. H. MacDonald
Phys. Rev. B 78,(2011) 121401(R)
\bibitem{lozovik}
	 Y.E. Lozovik, V.I. Yudson,
%	 "`Novel mechanism of superconductivity – pairing of spatially separated electrons and holes"',
Sov. Phys. JETP,1976,44: 389-397\\
	S. I. Shevchenko 
	%SI "`Theory of superconductivity of systems with pairing of spatially separated electrons and holes"', 
	Sov. J. Low Temp. Phys.1976, 2: 251-256.
	
	\bibitem{Gorsky:2017sgy} 
  A.~Gorsky and F.~Popov,
%  ``On magnetic and vortical susceptibilities of the Cooper condensate,''
  Phys.\ Lett.\ B {\bf 774}, 135 (2017)
  doi:10.1016/j.physletb.2017.09.040
  [arXiv:1707.05142 [cond-mat.supr-con]].
\end{thebibliography}
\end{document}